\documentstyle[amstex,amssymb,aps,twocolumn]{revtex}

\begin{document}
\title{Realization of Universal Optimal Quantum Machines by Projective Operators
and Stochastic Maps}
\author{F. Sciarrino, C. Sias, M. Ricci and F.\ De Martini}
\address{Istituto Nazionale per la Fisica della Materia, Dipartimento di Fisica,\\
Universit\`{a} di Roma ''La Sapienza'', Roma, 00185 - Italy}
\date{\today}
\maketitle

\begin{abstract}
Optimal quantum machines can be implemented by linear projective operations.
In the present work a general qubit symmetrization theory is presented by
investigating the close links to the qubit purification process and to the
programmable teleportation of any generic optimal anti-unitary map. In
addition, the contextual realization of the $N\rightarrow M$ cloning map and
of the teleportation of the $N\rightarrow \left( M-N\right) $ universal NOT
gate is analyzed by a novel and very general angular momentum theory. An
extended set of experimental realizations by state symmetrization linear
optical procedures is reported. These include the $1\rightarrow 2$ cloning
process, the UNOT\ gate and the quantum tomographic characterization of the
optimal partial transpose map of polarization encoded qubits.
\end{abstract}


\section{Introduction}

At a fundamental level quantum information (QI)\ consists of the set of
rules that identify and characterize the physical transformations that are
applicable to the quantum state of any information system. Because of the
constraints established by the quantum rules it is found that several
classical information tasks are\ forbidden or cannot be perfectly extended
to the quantum world. A well known and relevant QI\ limitation consists of
the impossibility of perfectly cloning (copying) any unknown qubit $\left|
\phi \right\rangle $ \cite{Woot82}. In other words, the map $\left| \phi
\right\rangle \stackrel{U}{\rightarrow }\left| \phi \right\rangle \left|
\phi \right\rangle $ cannot be realized by Nature because it does not belong
to the set of Completely Positive $(CP)$\ maps, i.e. the only ones
consistent with all requirements of quantum mechanics \cite{Kraus}. This
basic result may be the most fundamental difference between classical and
quantum information theory. Another forbidden operation is the NOT gate that
maps any $\left| \phi \right\rangle $ in its orthogonal state $\left| \phi
^{\perp }\right\rangle $ \cite{Bech99}. Even if these two processes are
unrealizable in their exact forms, they can be optimally approximated by the
so-called {\it universal quantum machines}, i.e. which exhibit the minimum
possible noise. A better understanding of these devices is important since
the exact characterization of the quantum constraints within basically
simple QI\ processes is useful to design more sophisticated algorithms and
protocols and to assess the limit performance of complex networks, such as a
quantum computer. In this paper we shall analyze the {\it optimal}
realizations of the NOT gate and of the cloning machine within the
enlightening perspective suggested by the new linear optical method that has
been recently adopted to achieve such realizations.

The $efficiency$ \ of a gate, i.e. that measures how close its action is to
the desired one, is generally quantified by the {\it fidelity} ${\cal F}$. $%
{\cal F}=1$ implies a perfect implementation, while noisy processes
correspond to ${\cal F}<1$. The Universal NOT (UNOT) gate, the optimal
approximation of the NOT gate, maps $N$ identical input qubits $\left| \phi
\right\rangle $ into $M$ optimal flipped ones in the state $\sigma _{out}$.
It achieves the fidelity : ${\cal F}_{N\rightarrow M}^{\ast }(\left| \phi
^{\perp }\right\rangle ,\sigma _{out})$ = $\langle \phi ^{\perp }|\sigma
_{out}|\phi ^{\perp }\rangle $ = $(N+1)/(N+2)$ that depends only on the
number of the input qubits \cite{Gisi99}. Indeed the fidelity of the UNOT
gate is exactly the same as the optimal quantum estimation fidelity \cite
{Mass95}. This\ means that such process may be modelled as a ''classical'',
i.e. exact, preparation of $M$ identical flipped qubits following the
quantum, i.e. inexact, estimation of $N$ input states. Only this last
operation is affected by noise. Only in the limit $N\rightarrow \infty $ a
perfect estimation of the input state is achieved and a perfect flipping
operation is also realized. Differently from the UNOT gate, the Universal
Optimal Quantum Cloning Machine (UOQCM), which transforms $N$ identical
qubits $\left| \phi \right\rangle $ into $M$ identical copies $\rho _{out}$,
achieves as optimal fidelity : ${\cal F}_{N\rightarrow M}(\left| \phi
\right\rangle ,\rho _{out})$ = $\langle \phi |\rho _{out}\left| \phi
\right\rangle $ = $(NM+M+N)/(MN+2M)$ = $(N+1+\beta )/(N+2)$ with $\beta
\equiv N/M\leq 1\;$\cite{Gisi97,Brus98,Buze98}. As we can see ${\cal F}%
_{N\rightarrow M}(\left| \phi \right\rangle ,\rho _{out})$ is larger than
the one obtained by the $N$ estimation approach and reduces to that result
for $\beta \rightarrow 0$, i.e. for an infinite number of copies. Of course
the zero-cloning condition is expressed by $\beta =1$ and ${\cal F}%
_{N\rightarrow N}=1$. The extra positive term $\beta $ in the above
expression accounts for the excess of quantum information which, originally
stored in $N$ states, is optimally redistributed by entanglement among \ the 
$M-N$ remaining blank qubits encoded by UOQCM \cite{Buze96}. Precisely, the
entanglement is established by the cloning process between the blank qubits
and the machine itself which may be modelled as a ''ancilla'' information
system.

The first conceptual approach to the realization of these transformations is
based on finding a suitable unitary operator $U_{NM}$, deterministically
realized by means of a quantum network and acting on $N$ input qubits and on 
$2(M-N)$ ancillary qubits. At the output of this device we obtain $M$ and $%
(M-N)$ qubits which are, respectively, the optimal clones and the best
flipped qubits of the input ones \cite{Buze97}. A different approach to the
probabilistic implementation of the $N\rightarrow M$ cloning process has
been proposed by Werner \cite{Wern98}. It is based on the action of a
projective operation on the symmetric subspace of the $N$ input qubits and $%
(M-N)$ blank ancillas. This transformation assures the uniform distribution
of the initial information into the overall system and guarantees that all
output qubits are indistinguishable. While previous realizations \cite
{DeMa02,Lama02,Pell03,Fase02,Cumm02} were inspired by the first approach,
the work reported in the present paper follows the path established by the
last theoretical proposals, as we see shortly.

In Quantum Optics let the qubit to be codified by the polarization state of
a single photon. Precisely in this context it was proposed to realize the $%
U_{NM}$ transformation by exploiting any amplification process e.g. realized
by the Quantum Injected Optical Parametric Amplifier (QIOPA) in the
entangled configuration \cite{DeMa98,Simo00}. Indeed the experimental
demonstrations of the UOQCM and the UNOT gate have been reported by
exploiting precisely this technique \cite{DeMa02,Lama02,Pell03}. The cloning
process (but not the UNOT gate) was also realized by a simple laser
amplifier in a $Er^{3+}-doped$ optical fiber \cite{Fase02}.

In the present work the more direct qubit symmetrization path proposed by
Werner was taken, as said. In this perspective an entirely new scenario has
been disclosed by the recent discovery that it is possible to implement
contextually the $1\rightarrow 2$ universal quantum cloning machine (UOQCM)
and the $1\rightarrow 1$ universal NOT gate by slightly modifying the
Quantum State Teleportation $(QST)$ protocol \cite{Ricc03}. Since in this
case the UNOT gate is transferred, i.e{\it . teleported}, in a different
location, it will be referred to as the {\it Tele-UNOT gate. }This indeed
realizes a novel QI\ protocol: the{\it \ ''teleportation of a quantum
operation''}.

In Section II of the present work the Tele-UNOT\ protocol is investigated
theoretically by a quantum network approach on the basis of the qubit
symmetrization process. Furthermore, this same process is shown to lead very
naturally to the efficient qubit purification protocol recently realized
experimentally \cite{Cira99,Ricc04}. Section III reports a most general
application of the symmetrization protocol, indeed one of the key results of
the present work:\ the programmable optimal teleportation of {\it any}
anti-unitary gate acting on qubits. In Section IV a general and
comprehensive theory of the $N\rightarrow M$ UOQCM and\ of the $N\rightarrow
(M-N)$ Tele-UNOT\ gate is given by adopting a novel and straightforward
theoretical approach to the problem, i.e. the well established$\ \ \left|
J,J_{z}\right\rangle $ angular momentum formalism of a general $J-$spin
system.\ A detailed account of the experimental realization of UOQCM and
Tele-UNOT\ with polarization encoded qubits by two alternative approaches is
reported in Section V \cite{Ricc03,Scia03}. In Section VI the experimental
implementation and characterization of the optimal transpose of a $2\times 2$
density operator by a stochastic method is reported. This is done in order
to investigate, at a deeper level how the UNOT gate can be realized starting
from the teleportation protocol. Finally, in Section VII the main results of
the work are summarized and considered in the perspectives of modern quantum
information, computation and estimation theories.

\section{Qubit Symmetrization}

The protocol that realizes the $1\rightarrow 2$ UOQCM and $1\rightarrow 1$
Tele-UNOT gate, involves two distant partners:\ Alice $({\cal A}{\Bbb )}$
and Bob $({\cal B}{\cal )}$. ${\cal A}$ holds the unknown input qubit $S$ in
a generic state $\left| \phi \right\rangle _{S}$, while ${\cal B}$ shall
finally receive this qubit encoded {\it optimally} by the UNOT
transformation of $\left| \phi \right\rangle _{S}$. Let ${\cal A}$ and $%
{\cal B}$ share the entangled singlet state of two qubits $A$, $B$: $\left|
\Psi ^{-}\right\rangle _{AB}=2^{-%
{\frac12}%
}\left( \left| \phi \right\rangle _{A}\left| \phi ^{\perp }\right\rangle
_{B}-\left| \phi ^{\perp }\right\rangle _{A}\left| \phi \right\rangle
_{B}\right) $, as in the QST protocol \cite{Benn93}. The adoption of \ the 
{\it singlet state} throughout this work guarantees, in virtue of its SU(2)
invariance, the universality of the overall process. The overall state of
the system $\left| \Omega _{SAB}\right\rangle $ reads: 
\begin{equation}
\left| \Omega _{SAB}\right\rangle =2^{-%
{\frac12}%
}\left| \phi \right\rangle _{S}\left( \left| \phi \right\rangle _{A}\left|
\phi ^{\perp }\right\rangle _{B}-\left| \phi ^{\perp }\right\rangle
_{A}\left| \phi \right\rangle _{B}\right)
\end{equation}
Let ${\cal A}$ to apply to the overall initial state $\left| \Omega
_{SAB}\right\rangle $ the projective operator $P_{SA}$ over the symmetric
subspace of the qubits $S$ and $A$: 
\begin{equation}
P_{SA}=({\Bbb {I}}_{SA}-\left| \Psi ^{-}\right\rangle _{SA}\left\langle \Psi
^{-}\right| _{SA})  \label{proiettore}
\end{equation}
The projection is successfully realized with probability $p=%
{\frac34}%
.$ In this case the normalized output state is 
\begin{gather}
\text{$\left| \Xi _{SAB}\right\rangle ${\sc =}}  \label{statoout} \\
=\text{{\sc $\sqrt{\frac{2}{3}}$$\left| \phi \right\rangle _{S}\left| \phi
\right\rangle _{A}\left| \phi ^{\perp }\right\rangle _{B}$-$\sqrt{\frac{1}{6}%
}$}{\sc ($\left| \phi \right\rangle _{S}\left| \phi ^{\perp }\right\rangle
_{A}$+$\left| \phi ^{\perp }\right\rangle _{S}\left| \phi \right\rangle _{A}$%
)}$\left| \phi \right\rangle _{B}$}  \nonumber
\end{gather}
A {\it one bit} of classical communications sent by ${\cal A}$ announces to $%
{\cal B}$ the success of the symmetrization protocol. This one leaves the
two qubits $S$, $A$ held by ${\cal A}$ in the same {\it mixed state} 
\begin{equation}
\rho _{S}=\rho _{A}=\frac{5}{6}\left| \phi \right\rangle \left\langle \phi
\right| +\frac{1}{6}\left| \phi ^{\perp }\right\rangle \left\langle \phi
^{\perp }\right|  \label{cloningmatrix}
\end{equation}
which represent the {\it optimal} output of the $1\rightarrow 2$ cloning
process for the input state $\left| \phi \right\rangle _{S}$ with the
expected fidelity : ${\cal F}_{1\rightarrow 2}=\frac{5}{6}$. Contextually,
the qubit $B$ held by ${\cal B}$ is left by the protocol in the mixed state 
\begin{equation}
\rho _{B}=\frac{1}{3}\left| \phi \right\rangle \left\langle \phi \right| +%
\frac{2}{3}\left| \phi ^{\perp }\right\rangle \left\langle \phi ^{\perp
}\right|
\end{equation}
and again the corresponding fidelity of the $1\rightarrow 1$ optimal
Tele-NOT process is the expected one:$\ {\cal F}_{1\rightarrow 1}^{\ast }=%
\frac{2}{3}$.

Note that the presence of the entangled state $\left| \Psi ^{-}\right\rangle
_{AB}\ $is not strictly necessary for the sole implementation of the quantum
cloning process as, for this purpose we could apply $P_{SA}$ to an initial
state $\left| \phi \right\rangle _{S}\left\langle \phi \right| _{S}\otimes 
\frac{I_{A}}{2}$ as shown experimentally by \cite{Scia03}.

\subsection{Implementation of the projection by a quantum network}

The projector $P_{SA}$ (\ref{proiettore}) into the symmetric space can be
implemented by means of a quantum circuit, in analogy with the protocol
devised for QST \cite{Scia03}. The projection is obtained by combining
Hadamard gates, $C-NOT$ gates, a Toffoli gate and the projective measurement
of an ancilla qubit $\widetilde{a}$, initially in the state $\left|
0\right\rangle _{\widetilde{a}}$ (Fig.2). Let us analyze here in more
details the logic of this network. The box ''EPR preparation'' prepares the
singlet state $\left| \Psi ^{-}\right\rangle _{AB}$ starting from the qubits 
$\left| 1\right\rangle _{A}$ and $\left| 1\right\rangle _{B}$. Hence the
state of the overall system is 
\begin{gather}
\left| \phi \right\rangle _{S}\otimes \left| \Psi ^{-}\right\rangle
_{AB}\otimes \left| 0\right\rangle _{\widetilde{a}}= \\
=\frac{1}{2}\left[ 
\begin{array}{c}
{\small -}\left| \Psi ^{-}\right\rangle _{SA}\left| \phi \right\rangle _{B}%
{\small -}\left| \Psi ^{+}\right\rangle _{SA}{\small \sigma }_{Z}\left| \phi
\right\rangle _{B}{\small +} \\ 
\left| \Phi ^{-}\right\rangle _{SA}{\small \sigma }_{X}\left| \phi
\right\rangle _{B}{\small +}\left| \Phi ^{+}\right\rangle _{SA}{\small %
\sigma }_{Z}{\small \sigma }_{X}\left| \phi \right\rangle _{B}
\end{array}
\right] \otimes \left| 0\right\rangle _{\widetilde{a}}  \nonumber
\end{gather}
The box labelled {\bf (1)} transforms the state $\left| \Psi
^{-}\right\rangle _{SA}$ into $\left| 1\right\rangle _{S}\left|
1\right\rangle _{A},$ while the other three Bell states $\left\{ \left| \Psi
^{+}\right\rangle _{SA},\left| \Phi ^{-}\right\rangle _{SA},\left| \Phi
^{+}\right\rangle _{SA}\right\} $ are respectively transformed into $\left\{
\left| 0\right\rangle _{S}\left| 1\right\rangle _{A},\left| 1\right\rangle
_{S}\left| 0\right\rangle _{A},\left| 0\right\rangle _{S}\left|
0\right\rangle _{A}\right\} .$ By means of a Toffoli gate \cite{Pres}, the
state $\left| 1\right\rangle _{S}\left| 1\right\rangle _{A}$ induces the
flipping of the qubit $\widetilde{a}$ from $\left| 0\right\rangle _{%
\widetilde{a}}$ to $\left| 1\right\rangle _{\widetilde{a}},$ whereas the
other states leave the qubit $\widetilde{a}$ unaltered. The state after the
Toffoli gate operation reads 
\begin{equation}
\text{$\frac{1}{2}\left[ 
\begin{array}{c}
\text{{\sc -$\left| 1\right\rangle _{S}\left| 1\right\rangle _{A}\left| \phi
\right\rangle _{B}\left| 1\right\rangle _{\widetilde{a}}$-$\left|
0\right\rangle _{S}\left| 1\right\rangle _{A}\sigma _{Z}\left| \phi
\right\rangle _{B}\left| 0\right\rangle _{\widetilde{a}}$+}} \\ 
\text{{\sc $\left| 1\right\rangle _{S}\left| 0\right\rangle _{A}\sigma
_{X}\left| \phi \right\rangle _{B}\left| 0\right\rangle _{\widetilde{a}}$+$%
\left| 0\right\rangle _{S}\left| 0\right\rangle _{A}\sigma _{Z}\sigma
_{X}\left| \phi \right\rangle _{B}\left| 0\right\rangle _{\widetilde{a}}$}}
\end{array}
\right] $}
\end{equation}
Finally the action of the box labelled {\bf (2) }restores the initial states
of the qubits $S$ and $A$ leading to$\left| \Sigma \right\rangle _{SAB%
\widetilde{a}}$ equal to: 
\begin{equation}
\frac{1}{2}\left[ 
\begin{array}{c}
{\small -}\left| \Psi ^{-}\right\rangle _{SA}\left| \phi \right\rangle
_{B}\left| 1\right\rangle _{\widetilde{a}}{\small -}\left| \Psi
^{+}\right\rangle _{SA}{\small \sigma }_{Z}\left| \phi \right\rangle
_{B}\left| 0\right\rangle _{\widetilde{a}}{\small +} \\ 
\left| \Phi ^{-}\right\rangle _{SA}{\small \sigma }_{X}\left| \phi
\right\rangle _{B}\left| 0\right\rangle _{\widetilde{a}}{\small +}\left|
\Phi ^{+}\right\rangle _{SA}{\small \sigma }_{Z}{\small \sigma }_{X}\left|
\phi \right\rangle _{B}\left| 0\right\rangle _{\widetilde{a}}
\end{array}
\right] 
\end{equation}
If the projective measurement on the ancilla qubit $\widetilde{a}$ gives as
result ''$1"$ the qubits $A$ and $S$ end up in the state $\left| \Psi
^{-}\right\rangle _{SA}$ while the qubit $B$ is in the state $\left| \phi
\right\rangle ,$ which has been then teleported from ${\cal A}$ to ${\cal B}.
$ If we obtain the result ''$0"$ the overall state becomes equal to $\left|
\Xi _{SAB}\right\rangle $\ (\ref{statoout}).\ The result of the ancilla
measurement is communicated to Bob and we realize the optimal quantum
cloning machine and the $Tele-UNOT$ gate of the input qubit $\left| \phi
\right\rangle $.

The circuit proposed by Brassard \cite{Bras96} to model QST, and then
realized by NMR techniques \cite{Niel98} achieves teleportation by means of
single qubit gates and $C-NOT$ gates. The present scheme somewhat retraces
the path of that circuit but there an ancilla state and a Toffoli gate
replace the Bell-measurement device with the detection of the realization of
state-symmetrization by the measurement apparatus at the site ${\cal A}$ .
Moreover, the circuit proposed by Bu\v{z}ek ${\em et}$ ${\em al}$. \cite
{Buze97} and realized adopting NMR technique \cite{Cumm02}, to model the
quantum cloning and the UNOT gate in the conventional devices, e.g. in a
QIOPA\ system \cite{Pell03}, differs from the present one since there QST is
not considered explicitly.

\subsection{Purification of single qubits.}

The circuit above represents a versatile tool for physically implementing
several relevant QI processes based on the state symmetrization process. For
instance it can be adopted to implement the optimal {\it quantum purification%
} of two qubits according to the symmetrization scheme proposed by Cirac 
{\it et al.} \cite{Cira99}. This one addresses the issue of the purification
of $N$ equally prepared qubits in the mixed state $\rho =\xi \left| \phi
\right\rangle \left\langle \phi \right| +{\frac{1}{2}}(1-\xi ){\Bbb I}$,
where $0\leq \xi \leq 1$. The procedure allows to distill out of a set of
mixed states a subset of states with a higher degree of purity, i.e. the
state purity is enhanced by filtering out some amount of the noise. The
purification scheme for $N=2$, consisting of a projection of two
polarization $(\pi )$ encoded qubits onto the symmetric subspace, can be
implemented by means of a symmetric beam-splitter $(BS)$, was recently
reported by \cite{Ricc04}.\ It can be easily checked that this purification
protocol can be modelled by the quantum network of Fig.2.

In order to further enlighten the connection between the purification and
cloning processes, let us consider the action of the symmetric projector on
two non entangled qubits having the same orientation on the Bloch sphere
but, generally, with different degree of mixedness: 
\begin{eqnarray}
\rho _{S} &=&\frac{1+\lambda _{S}}{2}\left| \phi \right\rangle \left\langle
\phi \right| +\frac{1-\lambda _{S}}{2}\left| \phi ^{\perp }\right\rangle
\left\langle \phi ^{\perp }\right| \\
\rho _{A} &=&\frac{1+\lambda _{A}}{2}\left| \phi \right\rangle \left\langle
\phi \right| +\frac{1-\lambda _{A}}{2}\left| \phi ^{\perp }\right\rangle
\left\langle \phi ^{\perp }\right|
\end{eqnarray}
We apply the projector $P_{SA}$ to the overall system $\rho _{S}\otimes \rho
_{A}$. The success probability of the procedure is equal to $p=%
{\frac14}%
(3+\lambda _{A}\ast \lambda _{S})$. Let us introduce a parameter $\Delta =%
{\frac12}%
(\lambda _{A}+\lambda _{S})$ quantifying the information over the input
qubits. The mean fidelity{\it \ }${\cal F}_{in}$ of the input qubits is
found: ${\cal F}_{in}\equiv 
{\frac12}%
\left( \left\langle \phi \right| \rho _{S}\left| \phi \right\rangle
+\left\langle \phi \right| \rho _{A}\left| \phi \right\rangle \right) =$ $%
{\frac12}%
(1+\Delta ).$ After the projection the output qubits, which are {\it equal}
since they belong to the symmetric subspace, are 
\begin{equation}
\rho _{out}=\frac{1+\lambda _{out}}{2}\left| \phi \right\rangle \left\langle
\phi \right| +\frac{1-\lambda _{out}}{2}\left| \phi ^{\perp }\right\rangle
\left\langle \phi ^{\perp }\right|
\end{equation}
with $\lambda _{out}=\frac{\Delta }{p}$. The fidelity of the output qubits
reads 
\begin{equation}
{\cal F}_{out}(\rho _{out},\left| \phi \right\rangle )=%
{\frac12}%
(1+\Delta p^{-1})
\end{equation}

For $\lambda _{S}=1$ and $\lambda _{A}=0,$ $p=\frac{3}{4},$ ${\cal F}%
_{in}=3/4$ and ${\cal F}_{out}=5/6.$ These values indeed correspond to the
optimal quantum cloning based on symmetrization \cite{Scia03}. For the case
of the qubit purification protocol, $\lambda _{S}=\lambda _{A}=\lambda $ the
following relations hold: $p=%
{\frac14}%
(3+\lambda ^{2})$, ${\cal F}_{in}=%
{\frac12}%
(1+\lambda )$ and ${\cal F}_{out}=%
{\frac12}%
(1+\lambda p^{-1})$.

\section{Programmable Optimal Teleportation of any Anti-unitary Map.}

As a consequence of the complete positivity (CP)\ character of any
realizable physical map \cite{Pres}, any anti-unitary transformation cannot
be implemented with fidelity ${\cal F}=1$. Any general anti-unitary operator 
$A$ can be expressed as $A=U^{A}K$, where $U^{A}$ is a unitary operator
depending on $A$ and $K$ is the complex-conjugate operator that transforms
any coefficient multiplying a ket standing at the right of $K$ into its
complex conjugate\cite{Saku96}. $K\;$implements the transpose map of the
density matrix $\rho $: ${\cal E}_{trans}\left( \rho \right) \equiv \rho ^{T}
$. For instance the NOT operator is: $A^{NOT}$ $=\sigma _{Y}K$, where $%
\sigma _{Y}$ is a Pauli operator. We can then express $A$ as: 
\begin{equation}
A=U^{A}\sigma _{Y}A^{NOT}  \label{sakurai}
\end{equation}
Precisely, let to express the given ''impossible'' anti-unitary processes
as: $A\left| \phi \right\rangle =\left| \phi ^{A}\right\rangle $ and $%
A^{NOT}\left| \phi \right\rangle =\left| \phi ^{\perp }\right\rangle $. In
the previous Sections we have seen how to implement the optimal
approximation of the NOT operator with fidelity ${\cal F}^{\ast }=\frac{2}{3}
$. Such transformation is performed by the map: ${\cal E}_{UNOT}\left(
\left| \phi \right\rangle \left\langle \phi \right| \right) $= $\frac{2}{3}%
\left| \phi ^{\perp }\right\rangle \left\langle \phi ^{\perp }\right| +\frac{%
1}{3}\left| \phi \right\rangle \left\langle \phi \right| $. We may now ask
what is the value of ${\cal F}^{\ast }$ of the optimal approximation to the
general anti-unitary operator $A$. It is easy to show that such value is
again ${\cal F}_{A}=\langle \phi ^{A}|\rho _{out}\left| \phi
^{A}\right\rangle =\frac{2}{3}$, the same as for the U-NOT one. Indeed,
consider the action over the input density matrix $\rho =\left| \phi
\right\rangle \left\langle \phi \right| $ of the map ${\cal E}_{A}(\rho )$.
In virtue of Eq. (\ref{sakurai}) is $A(A^{NOT})^{-1}=U^{A}\sigma _{Y}$ and
then 
\begin{gather}
{\cal E}_{A}\left( \rho \right) =U^{A}\sigma _{Y}{\cal E}_{UNOT}\left( \rho
\right) \sigma _{Y}U^{A+}=  \label{optimalantiunitary} \\
=\frac{2}{3}\left| \phi ^{A}\right\rangle \left\langle \phi ^{A}\right| +%
\frac{1}{3}\left| \phi ^{A\perp }\right\rangle \left\langle \phi ^{A\perp
}\right|   \nonumber
\end{gather}
leading to ${\cal F}_{A}=\frac{2}{3}$. If we could approximates $A$ with a
higher fidelity, then we could also implement a UNOT gate with fidelity $%
{\cal F}>\frac{2}{3}$, which is impossible. We can then assert that the
maximum fidelity achievable in a optimal universal approximation\ to any
general anti-unitary transformation\ applied to one qubit is ${\cal F}=\frac{%
2}{3}$. In other words, this last ${\cal F}$ value may be thought to
establish a $class$ of one qubits anti-unitary maps.

Let us generalize the above concepts in the framework of the QST protocol.
Note first that, by exploiting the result of Eq.(\ref{optimalantiunitary})
any {\it optimal anti-unitary operation} can be teleported by adopting a 
{\it different} entangled state in the protocol described in Section II. Let
Alice and Bob share the maximally entangled state obtained by any {\it local 
}transformation of the {\it singlet}: $\left| \Psi \right\rangle
_{AB}=\left( {\Bbb I}_{A}{\Bbb \otimes }U_{B}^{\dagger }\right) \left| \Psi
^{-}\right\rangle _{AB}$ =$\left( U_{A}{\Bbb \otimes I}_{B}\right) \left|
\Psi ^{-}\right\rangle _{AB}$ where $U=exp(-i\phi {\bf \ \sigma }\bullet 
{\bf n)/}2$ =$[\cos (\phi /2{\bf )}{\Bbb I}$ ${\bf -\ }i\sin (\phi /2{\bf %
)\sigma \bullet n]}$ is a general unitary that can be applied either at the
Alice's or at the Bob's sites. Let us discuss here the first option, the
most interesting one, by referring again to the $Tele-UNOT$\ protocol \cite
{Ricc03}. After projection into the symmetric subspace, Alice detects the
two optimal clones of the input qubit $\rho _{S}$ while, conditionally, Bob
detects the qubit $\rho _{B}=U^{\dagger }{\cal E}_{UNOT}\left( \rho
_{S}\right) U$, i.e. resulting from the application to $\rho _{S}$ of the
optimal approximation of the anti-unitary\ operator $U^{\dagger }\sigma
_{Y}K $. The choice of $U$, applied in the Alice's site (or, alternatively
the one of $U^{\dagger }$ applied at Bob's site) establishes the class of
all anti-unitary operators $A$ to be teleported with fidelity ${\cal F}_{A}=%
\frac{2}{3}$. For instance, according to the discussion above, $U=$ ${\Bbb I}
$, i.e. $\phi =0$, leads to the optimal $Tele-UNOT$ gate, while $U=$ ${\Bbb %
\sigma }_{Y}$ , i.e. $\phi =\pi /2$ and ${\bf \sigma }\bullet {\bf n=}{\Bbb %
\sigma }_{Y}$, leads to the optimal $Tele-Transposition$ gate \cite{Busc03},
etc. It is quite remarkable that, according to the QST\ concept, both the
input state $\rho _{S}$ and the operator $U$\ are realized and kept under
control at the Alice's site while the optimal anti-unitary gate is
transferred far apart by the noiseless non-local channel. This novel
universal and optimal ''{\it programmable gate-teleportation''} process is
represented in Fig. 1 by the insertion in the logical circuit of the general
unitary $U$. It may considered as a relevant realization of the universal
quantum processor reported by \cite{Hill02}.

\section{Angular momentum theory of the $N\rightarrow M$ Universal Optimal
Cloning and of the $N\rightarrow (M-N)$ U-NOT gate.}

So far the quantum cloning of $N^{\prime }=1$ qubit into $M^{\prime }=2$
qubits and the optimal flipping of $N^{\prime }=1$ qubit into $(M^{\prime
}-N^{\prime })=1$ qubit have been considered. By generalizing the Tele-UNOT
protocol to $N>1$ identical input qubits and $(M-N)>1$ entangled pairs we
obtain the teleportation of $(M-N)$ qubits optimally flipped at Bob's site
and, contextually, the optimal realization at Alice's site of the $%
N\rightarrow M$ cloning. Let us briefly describe this protocol: $N$ input
qubits in the state $\left| \phi \right\rangle =\alpha \left| 0\right\rangle
+\beta \left| 1\right\rangle =U_{\phi }\left| 0\right\rangle $\ are sent to
Alice who shares with Bob $(M-N)$ entangled pairs, all in the {\it singlet}
state to guarantee the universality of the process: $\left| \Psi
^{-}\right\rangle =2^{-1/2}(\left| 0\right\rangle \left| 1\right\rangle
-\left| 1\right\rangle \left| 0\right\rangle )$. The initial state of the
overall system is $\left| \Omega \right\rangle =\left| \phi \right\rangle
^{\otimes N}\left| \Psi ^{-}\right\rangle ^{\otimes (M-N)}=U_{\phi
}^{\otimes N}\left| 0\right\rangle ^{\otimes N}\left| \Psi ^{-}\right\rangle
^{\otimes (M-N)}$. Alice applies the projector $P_{sym}^{M}$ over the
symmetric subspace to her $M$ qubits, i.e. $N$ input qubits + $(M-N)$
ancilla qubits, and communicates to Bob the positive realization of the
symmetrization procedure by means of one classical bit. The overall
input-output protocol is enlightened by the scheme reported in Fig. 3.

In order to simplify the demonstration of the optimality, we exploit the
universality of the projection procedure. Indeed: 
\begin{gather}
(P_{sym}^{M}\otimes {\Bbb I}_{B})[U_{\phi }^{\otimes N}\left| 0\right\rangle
^{\otimes N}\left| \Psi ^{-}\right\rangle ^{\otimes (M-N)}]=
\label{covariance} \\
=U_{\phi }^{\otimes (2M-N)}P_{sym}^{M}\otimes {\Bbb I}_{B}\left|
0\right\rangle ^{\otimes N}\left| \Psi ^{-}\right\rangle ^{\otimes (M-N)} 
\nonumber
\end{gather}
for any $U_{\phi }\in SU(2).$ ${\Bbb I}_{B}$ is the identity operator acting
on the Hilbert space of Bob's qubits. The covariance property expressed in (%
\ref{covariance}) is assured by the invariance of the singlet state for
simultaneous unitary operations on the two qubits $\left| \Psi
^{-}\right\rangle =U_{\phi }^{\otimes 2}\left| \Psi ^{-}\right\rangle $ and
by the commutation property of the projector $P_{sym}^{M}$: $\left[
P_{sym}^{M},U^{\otimes M}\right] =0$. The covariance property allows us to
assume as input state $\left| \phi \right\rangle =\left| 0\right\rangle $
without lack of generality.

In the following part of this Section a very general and comprehensive
theory of the universal optimal cloning and U-NOT\ gate is given by
adopting, in a straightforward fashion,\ the well established$\ \ \left|
J,J_{z}\right\rangle $ angular momentum formalism of a general $J-$spin
system \cite{Edmonds}. The overall\ symmetric state of the $N$ input qubits
is taken to corresponds to a system with total spin $\frac{N}{2}$: $\left|
0\right\rangle ^{\otimes N}\circeq \left| \frac{1}{2},\frac{1}{2}%
\right\rangle ^{\otimes N}=\left| \frac{N}{2};\frac{N}{2}\right\rangle $.
Accordingly, the joint state of the $N$ input qubits and of $(M-N)$
entangled pairs $\left| \Omega \right\rangle $ is re-expressed in the spin
formalism as $\left| \frac{N}{2};\frac{N}{2}\right\rangle \otimes \left|
0,0\right\rangle ^{\otimes M-N}=\left| \frac{N}{2};\frac{N}{2}\right\rangle $
because the $M-N$ singlets contribute to the total spin with $J=J_{z}=0$.
The symmetrization projector $P_{sym}^{M}$ is defined as: $%
P_{sym}^{M}=\sum_{k=0}^{M}\left| \frac{M}{2};\frac{M}{2}-k\right\rangle
\left\langle \frac{M}{2};\frac{M}{2}-k\right| $because all symmetrized
''cloned'' spins are equally directed in the Poincare' space and the
projection is over the maximum allowed value of $J$. After the action of $%
P_{sym}^{M}$ we obtain the following normalized output state 
\begin{gather}
\left| \Omega ^{\prime }\right\rangle =\frac{P_{sym}^{M}\otimes {\Bbb I}%
_{B}\left| \frac{N}{2};\frac{N}{2}\right\rangle }{\left| P_{sym}^{M}\otimes 
{\Bbb I}_{B}\left| \frac{N}{2};\frac{N}{2}\right\rangle \right| }=
\label{outputNM} \\
\sum_{k=0}^{M-N}{\small b}_{k}\left| \frac{M}{2};\frac{M}{2}-k\right\rangle
_{A}{\small \otimes }\left| \frac{M-N}{2};\frac{-(M-N)}{2}+k\right\rangle
_{B}  \nonumber
\end{gather}
where$\ b_{k}=\left( -1\right) ^{k}\sqrt{\frac{N+1}{M+1}}\sqrt{\frac{%
(M-N)!(M-k)!}{M!(M-N-k)!}}$ is the Clebsch - Gordan coefficient $\langle
j_{1};m_{1k};j_{2};m_{2k}\left| j_{1};j_{2};j_{TOT};m_{TOT}\right\rangle $
with $j_{1}=\frac{M}{2}$, $j_{2}=\frac{M-N}{2}$, $m_{1k}=\frac{M}{2}-k$, $%
m_{2k}=\frac{-(M-N)}{2}+k$, $j_{TOT}=\frac{N}{2}$, $m_{TOT}=\frac{N}{2}$ ( 
\cite{Edmonds}, Ch.3.6). In the above representation, the overall output
state of the cloner is written as the composition of two angular momenta: $%
{\bf J}_{C},{\bf J}_{AC}$\ defined respectively over the ''cloning'' and
''anticloning'' output channels. In the present context, these angular
momenta correspond to the output states realized at the Alice's and Bob's
sites respectively. The success probability of the procedure is $\left|
P_{sym}^{M}\otimes {\Bbb I}_{B}\left| \frac{N}{2};\frac{N}{2}\right\rangle
\right| ^{2}\;$= $\frac{1}{2^{M-N}}\frac{1+M}{1+N}$. We note that the $(M-N)$
Bob's qubits assume the maximum allowed value of $J=\frac{M-N}{2}$, thus
they lie in the symmetric subspace in analogy with the Alice's ones.

The fidelities of the cloning and of the UNOT processes can be inferred
re-arranging the output state (\ref{outputNM}) as follow 
\begin{gather}
\left| \Omega ^{\prime }\right\rangle {\small =} \\
\sum_{k=0}^{M-N}{\small b}_{k}\left| \left\{ (M-k)\phi ;k\phi ^{\perp
}\right\} \right\rangle _{A}{\small \otimes }\left| \left\{ k\phi ;\left(
M-N-k\right) \phi ^{\perp }\right\} \right\rangle _{B}  \nonumber
\end{gather}
The notation $\left| \left\{ p\phi ;q\phi ^{\perp }\right\} \right\rangle $
stands for a total symmetric combination of $p$ qubits in the state $\left|
\phi \right\rangle $ and of $q$ qubits in the state $\left| \phi ^{\perp
}\right\rangle .$ All the $\left( p+q\right) $ qubits belonging in such
state have an identical reduced density matrix equal to 
\begin{equation}
\rho _{p,q}=\frac{p}{p+q}\left| \phi \right\rangle \left\langle \phi \right|
+\frac{q}{p+q}\left| \phi ^{\perp }\right\rangle \left\langle \phi ^{\perp
}\right|   \label{reducedqubits}
\end{equation}
The fidelity ${\cal F}_{CLON}$ of the cloning process is thus 
\begin{eqnarray}
{\cal F}_{N\rightarrow M} &=&\sum_{k=0}^{M-N}\left| b_{k}\right|
^{2}\left\langle \phi \right| \rho _{M-k,k}\left| \phi \right\rangle = 
\nonumber \\
&=&\sum_{k=0}^{M-N}\left| b_{k}\right| ^{2}{\cal F}_{CLON}^{k}=\frac{%
N+1+\beta }{N+2}
\end{eqnarray}
where $\beta =\frac{N}{M}$ and ${\cal F}_{CLON}^{k}=\frac{M-k}{M}$ is the
fidelity of the $k$-th term of the summation derived from the expression (%
\ref{reducedqubits}). The above expression of ${\cal F}_{N\rightarrow M}$
coincides with the one given in literature \cite{Buze96} and in Sect.I,
above. For the UNOT\ process we obtain 
\begin{eqnarray}
{\cal F}_{N\rightarrow (M-N)}^{\ast } &=&\sum_{k=0}^{M-N}\left| b_{k}\right|
^{2}\left\langle \phi ^{\perp }\right| \rho _{k,M-N-k}\left| \phi ^{\perp
}\right\rangle =  \nonumber \\
&=&\sum_{k=0}^{M-N}\left| b_{k}\right| ^{2}{\cal F}_{UNOT}^{k}=\frac{N+1}{N+2%
}
\end{eqnarray}
where ${\cal F}_{UNOT}^{k}=\frac{M-N-k}{M-N}$. This value coincides with the
optimal fidelity given by Gisin and Massar \cite{Gisi97} of any general
measurement of $N$ equal and unknown qubits. Thus we retrieve the result by
which, as far as fidelity is concerned, the U-NOT\ process is equivalent to
a quantum estimation measurement followed by a ''classical'' inversion of
the corresponding outcomes \cite{DeMa02}. However, as already noted in
Sect.I, this is not the case for quantum cloning where the extra term $%
\varpropto \beta $ in Eq.19 accounts for the residual information stored in
the entanglement of the output ''clones''.

In analogy with the $1\rightarrow 2$ UOQCM protocol analyzed in Sect.II, we
note once again that the entangled source is not strictly necessary in order
to achieve solely the $\ N\rightarrow M$ cloning process as for this purpose
only $\left( M-N\right) $ ancilla qubits in a fully mixed state are needed 
\cite{Wern98}. Furthermore, as a further generalization of the results of
Sect.III above, by starting with pairs bearing a different entanglement
structure it is possible to teleport a generic, optimal anti-unitary
transformation. For instance, the adoption of the {\it triplet state} $%
\left| \Phi ^{+}\right\rangle =2^{-1/2}\left( \left| 0\right\rangle \left|
0\right\rangle +\left| 1\right\rangle \left| 1\right\rangle \right) $ leads
to the quantum cloning machine given by Gisin and Massar \cite{Gisi97} by
which the $\left( M-N\right) $ qubits teleported to Bob represent the {\it %
optimally transposed} transformation of the input qubits. More about this
process is reported in Section VI, below.

\section{Experimental Optical Implementations.}

In the experiments reported in \cite{Ricc03} and \cite{Scia03}, the input
qubit was codified into the {\it polarization} $(\pi )$ state of a single
photon belonging to the mode $k_{S}$: $\left| \phi \right\rangle _{S}=\alpha
\left| H\right\rangle _{S}+\beta \left| V\right\rangle _{S}$, whereas an
entangled pair $\left| \Psi ^{-}\right\rangle _{AB}$ of photons $A$ and $B$,
was generated on the modes $k_{A}$ and $k_{B}$ by Spontaneous Parametric
Down Conversion (SPDC). The {\it projective operation} in the space $%
H=H_{S}\otimes H_{A}$ was realized by exploiting the linear superposition of
the modes $k_{S}$ and $k_{A}$ within a $50:50$ {\it beam-splitter, }$BS_{A}$
(Figs.4-6)$.$ This superposition allows a partial Bell measurement on the $%
BS_{A}$ output states which is needed to implement the cloning machine and
the Tele-UNOT gate. Consider the overall output state realized on the two
modes $k_{1}\,$and $k_{2}$ of $BS_{A}$ and expressed by a superposition of
the Bell states: $\left\{ \left| \Psi ^{-}\right\rangle _{SA},\left| \Psi
^{+}\right\rangle _{SA},\left| \Phi ^{-}\right\rangle _{SA},\left| \Phi
^{+}\right\rangle _{SA}\right\} $. The realization of the singlet $\left|
\Psi _{SA}^{-}\right\rangle $ is identified by the emission of\ one photon
on each output mode of $BS_{A}$, while the realization of the other three
Bell states implies the emission of 2 photons either on mode $k_{1}\,$or on
mode $k_{2}$. This Ou-Mandel interference\ process, expressing a {\it Bose
mode coalescence} (BMC) of the two photons over the same mode, was
experimentally identified by a coincidence event between two detectors
coupled to the output mode $k_{2}$ by means of an additional $50:50$
beam-splitter. The identical effect expected on mode $k_{1}$ was neglected,
for simplicity. As\ just shown, this condition assured the simultaneous
experimental realization of the UNOT and UOQCM processes, here detected by a 
{\it post-selection} technique.

Note that, while the present Tele-UNOT protocol with $\pi -$encoded qubits
is fully realizable by linear optical methods, the full implementation by
these methods of the Bell measurement in the standard QST protocol is
impossible \cite{Cals01}. Hyper-entanglement with additional degrees of
freedom \cite{Boschi98} or a network of C-NOT gates are required to perform
that task. In general, by a balanced beam splitter a POVM measurement is
able to distinguish between the symmetric and the asymmetric components of
the overall state of two qubits. Indeed, the projection into the symmetric
space lies at the core of the cloning process, as epitomized by the present
work \cite{Wern98,Ricc03,Scia03,Irvi04}. The cloning process was
investigated in two independent experiments which enlighten different
features of the protocol. The first experiment, involving only two photons
generated by the same SPDC\ process, demonstrated that an entangled state is
not necessary for the UOQCM implementation and achieves a fidelity close to
the limit value. The second one, which adopts photons belonging to
independent sources, experimentally demonstrated the overall process.

\subsection{Ou Mandel Cloning}

In order to clone the input qubit $S$, an entangled state of the qubits $A$
and\thinspace $B$ was not necessary as only a qubit $A$ in a fully mixed
state was needed \cite{Scia03}. Hence we carried out a first experiment
involving only two qubits. A pair of photons with equal wavelength\ (wl) $%
\lambda =532nm$ and coherence-time $\tau _{coh}=80fs$, was generated by a
SPDC process in a Type I BBO\ crystal in the initial polarization product
state $\left| H\right\rangle _{S}\left| H\right\rangle _{A}$ (Fig.4). The
non-linear (NL) crystal was pumped by a continuous-wave (cw) UV beam with wl 
$\lambda =266nm$, created by fourth-harmonics generation in a OPO cavity
(Coherent:MBD266) by a cw Nd:YAG laser (Coherent:VERDI) with wl $\lambda
=532nm$. This sophisticated system provided a true single mode UV beam with
linewidth 
\mbox{$<$}%
100 MHz and high power, up to 400 mW allowing a high coincidence rate. The
photons $S$ and $A$ from SPDC pairs were injected on the two input modes $%
k_{S}$ and $k_{A}$ of $BS_{A}$ with an adjustable mutual temporal delay $%
\Delta t$. The input qubit $\left| \phi \right\rangle _{S}$ associated with
mode $k_{S}$ was polarization encoded by means of a waveplate (wp) $WP_{S}$.
The transformation used to map the state $\left| H\right\rangle _{A}$ into $%
\rho _{A}=\frac{{\Bbb {I}_{A}}}{2}$ was achieved by stochastically rotating,
during each experimental run, a $\lambda /2$ wp $(WP_{A})$ inserted on the
mode $k_{A}$. In this way the statistical evolution of $\left|
H\right\rangle _{A}$ into two orthogonal states with equal probability was
achieved.

The $\pi -$state on the output mode $k_{2\text{ }}$of $BS_{A}$ was analyzed
by the combination of the wp $WP_{C}$ and of the {\it polarizing beam
splitter} (PBS): $PBS_{C}$. For each input $\pi -$state $\left| \phi
\right\rangle _{S}$, $WP_{C}$ was set in order to make $PBS_{C}$ to transmit 
$\left| \phi \right\rangle $ and reflect $\left| \phi ^{\perp }\right\rangle 
$. The ''cloned'' state $\left| \phi \phi \right\rangle $ could be detected
on mode $k_{2}$ by a two-photon counter, realized in our case by first
separating the two photons by an additional $50:50$ beam splitter $BS_{C}$
and then detecting the coincidence $[D_{C},D_{C}^{\prime }]$ between the
output detectors $D_{C}$ and $D_{C}^{\prime }$:\ Fig. 4. Any coincidence
between $D_{C}^{\ast }$ and $D_{C}$ corresponded to the realization of the
state $\left| \phi \phi ^{\perp }\right\rangle $. First consider the cloning
machine switched off by spoiling the interference of $S$ and $A$ in $BS_{A}$%
, i.e. by setting: $\Delta t=Z/(2c)>>\tau _{coh}$, being $Z$ a micrometrical
displacement of $BS_{A}$. In this case, since the states $\left| \phi \phi
\right\rangle $ and $\left| \phi \phi ^{\perp }\right\rangle $ were realized
with the same probability on mode $k_{2}$, the rate of coincidences detected
by $[D_{C},D_{C}^{\prime }]$ were expected to one half of the one detected
by $[D_{C}^{\ast },D_{C}^{\prime }]$. By turning on the cloning machine,
i.e. by setting $\Delta t\approx 0$ the output density matrices $\rho _{S},$ 
$\rho _{A}$ (\ref{cloningmatrix}) were realized on the mode $k_{2}$ with an
enhancement ratio $R=2$ of the counting rate by the set $[D_{C},D_{C}^{%
\prime }]$ and no rate enhancement by $[D_{C}^{\ast },D_{C}^{\prime }]$. All
adopted photodetectors $(D)$ were SPCM-200 single photon counters and
interferential filters with bandwidth (bwth) $\Delta \lambda =5nm$ were
placed behind them.

The experimental data are reported\ in Fig. 4 for three different input $\pi
-states$: $\left| \phi \right\rangle _{S}=\left| H\right\rangle $, $2^{-%
{\frac12}%
}(\left| H\right\rangle +\left| V\right\rangle )$, $2^{-%
{\frac12}%
}(\left| H\right\rangle +i\left| V\right\rangle )$. There square and
triangular marks refer respectively to the $[D_{C},D_{C}^{\prime }]$ and $%
[D_{C}^{\prime },D_{C}^{\ast }]$ coincidences versus the time setting $Z$.
The corresponding experimental values of the cloning fidelity{\it \ }${\cal F%
}=(2R+1)/(2R+2)$ are ${\cal F}_{H}=0.827\pm 0.002,$ ${\cal F}_{H+V}=0.825\pm
0.002,$ ${\cal F}_{H+iV}=0.826\pm 0.002$. These values are in good agreement
with the optimal value ${\cal F}_{1\rightarrow 2}=5/6=8.333$ corresponding
to the limit $S/N$ value: $R=2$. A similar experiment could be performed by
adopting a single photon source to produce an ancilla photon on mode $k_{2%
\text{ }}$ \cite{Sant02}.

\subsection{Cloning + Tele UNOT Gate}

To further investigate the UOQCM, in a second experiment we employed two
independent photons generated by uncorrelated processes: the input qubit was
obtained by strongly attenuating a coherent beam while the ancilla photon
was generated by a SPDC\ process. In order to observe the UOQCM process the
indistinguishability between the two photons at the output mode $k_{2\text{ }%
}$of $BS_{A}$ had to be attained by realizing a {\it single} output mode
condition with the best possible approximation, as we shall see shortly.

The source of the SPDC process was a Ti:Sa mode-locked pulsed laser
(Coherent: MIRA) with wl $\lambda =795nm$ and repetition rate $76MHz$ (Fig.
5). A weak beam, deflected from the laser beam by a partial reflecting
mirror $M$, was strongly attenuated by filters $(At)$ and delayed by $Z${\bf %
\ }$=2c\Delta t${\bf \ }via{\bf \ }a micrometrically adjustable optical
''trombone''. This beam was the source of the {\it quasi} single-photon
state injected into $BS_{A}$ over the mode $k_{S}$. The average number of
injected photons was $\overline{n}\simeq 0.1$. Different qubit states $%
\left| \phi \right\rangle _{S}$ were prepared via a $\lambda /2$ or $\lambda
/4$ wp $WP_{S}$. The UV\ laser beam with wl $\lambda _{p}=397.5$ $nm$,
generated by $2^{nd}-harmonics$ generation, excited the SPDC\ source of the 
{\it singlet} $\left| \Psi ^{-}\right\rangle _{AB}$. The photons $A$ and $B$
of each entangled pair were emitted over the modes $k_{A}$ and $k_{B}$ with
equal wls $\lambda =795nm$. All adopted photodetectors $(D)$ were SPCM-AQR14
single photon counters.

In order to observe the Ou-Mandel interference at the output of $BS_{A}$ a
high {\it spatial} indistinguishability of the photons $A$ and $S$ was
provided by a single mode selector $(MS),$ realized by a $5$ $m$ long
single-mode fiber, inserted on mode $k_{2}$ by a fiber coupler Thor Labs $%
KT110.$ The fixed $\pi -$transformation induced by the propagation inside
the fiber was compensated by a Babinet compensator $(BC)\;$and a $\lambda /2$
wp. The induced rotation was stable up to $1\%$ for more than 1 day. An
interferential filter $(IF)$ with bwth $\Delta \lambda =3nm$ placed in front
of each $D$ determined the coherence time of \ the detected pulses: $\tau
_{coh}\simeq 350fs$. Two fixed quartz plates $(Q)$ inserted on the modes $%
k_{A}$ and $k_{B}$ provided the compensation for the unwanted walk-off
effects due to the birefringence of the BBO ($\beta -$barium borate)
nonlinear $(NL)$\ crystal providing the source of SPDC.

In order to demonstrate the realization of the linear UOQCM\ process, the
states $\rho _{A}$ and $\rho _{S}$ of the clones $A$ and$\ S$ at the output
of $BS_{A}$ were investigated. According\ to the quantum analysis given in
Sect.II, \ expressed by Eq.(\ref{cloningmatrix}), we would expect $\rho
_{S}=\rho _{A}=(5\rho _{IN}+\rho _{IN}^{\perp })/6$ where $\rho _{IN}=\left|
\phi \right\rangle \left\langle \phi \right| $. The measurements were
realized on the $BS_{A}$ output mode $k_{2}$ by adopting the apparatus shown
in Fig. 5. The $\pi -state$ on this mode was analyzed by the combination of
the wp $WP_{C}$ and of the polarizer beam splitters $PBS$. For each input $%
\pi -state$ $\left| \phi \right\rangle _{S}$, $WP_{C}$ was set in order to
make $PBS$ to transmit $\left| \phi \right\rangle $ and reflect $\left| \phi
^{\perp }\right\rangle $. The ''cloned'' state $\left| \phi \phi
\right\rangle $ was detected on mode $k_{2}$ by the coincidence set $[D_{C}$ 
$D_{C}^{\prime }]$. The generation of an entangled pair was tested by
detection of one photon on the mode $k_{B}$ by $D_{B}$. Any coincidence
detected by the sets $[D_{C},D_{C}^{\prime },D_{B}]$ and $[D_{C},D_{C}^{\ast
},D_{B}]$ implied the realization of\ the states $\left| \phi \phi
\right\rangle $ and $\left| \phi \phi ^{\perp }\right\rangle $,
respectively. The experimental results of the {\it signal-to-noise} $(S/N)$
ratio $R$, carried out by coincidence measurements involving $%
[D_{C},D_{C}^{\prime },D_{B}]$ and $[D_{C},D_{C}^{\ast },D_{B}]$ are
reported\ in Fig. 5, again for the three different input $\pi -states$: $%
\left| \phi \right\rangle _{S}=\left| H\right\rangle $, $\left| \phi
\right\rangle _{S}=2^{-1/2}(\left| H\right\rangle +\left| V\right\rangle )$, 
$\left| \phi \right\rangle _{S}=2^{-1/2}(\left| H\right\rangle +i\left|
V\right\rangle )$. The square and triangular markers there refer
respectively to the $[D_{C},D_{C}^{\prime },D_{B}]$ and $[D_{C},D_{C}^{\ast
},D_{B}]$ coincidence plots vs the delay $Z.$ The following values of the 
{\it cloning fidelity }${\cal F}=(2R+1)/(2R+2)$ were found ${\cal F}%
_{H}=0.821\pm 0.003,$ ${\cal F}_{H+V}=0.813\pm 0.003,$ ${\cal F}%
_{H+iV}=0.812\pm 0.003$ to be compared with the {\it optimal} ${\cal F}%
_{1\rightarrow 2}=5/6\approx 0.833$ corresponding to the limit $S/N$ value: $%
R=2$. These results have been evaluated by taking into account the
reduction, by a factor $\xi =0.7$, of the $S/N$ ratio $R$ due to unwanted
coincidence rates attributable to the spurious simultaneous injection of two
photons on the mode $k_{S}$ and to simultaneous emission of two SPDC\ pairs.
The factor $\xi $ was carefully evaluated by a side experiment.

In order to realize the $Tele-UNOT$ protocol, the ''{\it Bose coalescence''}
\ process was detected on the output mode $k_{2}$ of $BS_{A}$ by the
coincidence $[D_{2}$ $D_{2}^{\ast }]$ as shown in Fig. 6. At Bob's site, the
polarization state on the mode $k_{B}$ was analyzed by the combination of
the wp $WP_{B}$ and of $PBS_{B}$. For each input $\pi -state$ $\left| \phi
\right\rangle _{S}$, $WP_{B}$ was set in order to make the $PBS_{B}$ to
transmit $\left| \phi \right\rangle _{B}$ and to reflect $\left| \phi
^{\perp }\right\rangle _{B}$, by then exciting $D_{B}$ and $D_{B}^{\ast }$
correspondingly. First consider the QST\ turned off, by setting the optical
delay $\left| Z\right| \gg c\tau _{coh}$. In this case, since the states $%
\left| \phi \right\rangle _{B}$ and $\left| \phi ^{\perp }\right\rangle _{B}$
were realized with the same probability on mode $k_{B}$, the rate of
coincidences detected by the $D-$sets $[D_{B},D_{2},D_{2}^{\ast }]$ and $%
[D_{B}^{\ast },D_{2},D_{2}^{\ast }]\ $were expected to be equal. By turning
on the QST, i.e. by setting $\left| Z\right| <<c\tau _{coh}$, the output
state $\rho _{B}^{out}=(2\rho _{IN}^{\perp }+\rho _{IN})/3$ was realized
then implying an {\it enhancement} by a factor $R=2$ of the counting rate $%
[D_{B}^{\ast },D_{2},D_{2}^{\ast }]$ and {\it no enhancement} of $%
[D_{B},D_{2},D_{2}^{\ast }]$. The corresponding 3-coincidence results shown
in Fig.6 and involving these 3-detector sets correspond to the injection of
three different input $\pi -states$: $\left| \phi \right\rangle _{S}=\left|
H\right\rangle $, $\left| \phi \right\rangle _{S}=2^{-1/2}(\left|
H\right\rangle +\left| V\right\rangle )$, $\left| \phi \right\rangle
_{S}=2^{-1/2}(\left| H\right\rangle +i\left| V\right\rangle )$. As such
these results indeed demonstrate the {\it universality} of the $Tele-UNOT$
process. In Fig. 6 the square and triangular markers refer respectively to
the $[D_{B}^{\ast },D_{2},D_{2}^{\ast }]$ and $[D_{B},D_{2},D_{2}^{\ast }]$
coincidences versus the delay $\Delta t.$ The $Tele-UNOT$ process was found
only to affect the $\left| \phi ^{\perp }\right\rangle _{B}$ component, as
expected. The $S/N$ ratio $R\;$was determined for each resonance curve as
the ratio between the values of the resonance peak, i.e. for $Z\simeq 0$,
and the no enhancement value, i.e. for $\left| Z\right| \approx 0$. The
experimental values of the UNOT\ {\it fidelity\ }${\cal F}^{\ast }=R/(R+1)$
were found, in correspondence with the three injected $\pi -states$ $\left|
\phi \right\rangle _{S}$: ${\cal F}_{H}=0.641\pm 0.005$, ${\cal F}%
_{H+V}=0.632\pm 0.006$, ${\cal F}_{H+iV}=0.619\pm 0.006$ to be compared with 
${\cal F}_{1\rightarrow 1}^{\ast }=R/(R+1)=2/3=0.667$. As for the cloning
experiment, the measured correcting factor $\xi =0.7$ has been used to
evaluate of the value of the fidelity.

\section{Stochastic Experimental Realization of the Optimal Transpose Map}

In Section II the link existing between all optimal anti-unitary operations
has been considered. In the present Section a simple and significant
implementation of the optimal approximation to the {\it transpose map }$K$
is reported by a different approach. The optimal transpose map ${\cal E}%
_{TR} $ has the following Kraus representation \cite{Pres} 
\begin{equation}
{\cal E}_{TR}(\rho )=\frac{1}{3}\left( {\Bbb I}\rho {\Bbb I}+\sigma _{X}\rho
\sigma _{X}+\sigma _{Z}\rho \sigma _{Z}\right)  \label{optimaltranspose}
\end{equation}
The action of the map ${\cal E}_{TR}$ can be viewed as the equiprobable
occurrence of three different operators, ${\Bbb I},\sigma _{X}$ and $\sigma
_{Z}$. Such transformation can be achieved either by the action of an
unitary operator into a larger system or by a stochastic evolution of the
system.

Let us outline the importance of the $K$ and ${\cal E}_{TR}(\rho )$
transformations in the context of quantum information. The transpose map $K$
is a $P-map$, as said. As such it transforms entangled states into
non-physical ones. It is exactly this property that makes the transposition
operation so important in all criteria of inseparability for two qubit
systems. A bipartite state $\rho _{AB}$ of two qubits, $A$ and $B$, is
entangled if and only if the density matrix ${\Bbb I}_{A}\otimes K_{B}(\rho
_{AB})$ has negative eigenvalues \cite{Horo96}, where the operation ${\Bbb I}%
_{A}\otimes K_{B}$ is commonly referred to {\it Partial Transpose operation}%
. An experimental limitation of this criteria is that it requires complete
knowledge of the state $\rho _{AB}$. Recently Horodecki and Ekert have found
an experimental method for a direct detection of quantum entanglement
exploiting the former criteria \cite{Horo02}. It consists of applying the
map $\widetilde{{\Bbb I}_{A}\otimes K_{B}}=\left[ \frac{1}{3}{\cal E}%
_{UNOT-A}\otimes {\cal E}_{DEP-B}+\frac{2}{3}{\Bbb I}_{A}\otimes {\cal E}%
_{TR-B}\right] $\ to the state $\rho _{AB}$ where ${\cal E}_{DEP}(\rho )=$\ $%
\frac{1}{4}\left( {\Bbb I}\rho {\Bbb I}+\sigma _{X}\rho \sigma _{X}+\sigma
_{Y}\rho \sigma _{Y}+\sigma _{Z}\rho \sigma _{Z}\right) $ represents a
depolarizing channel. Note, in the expression above, the appearance of both
the optimal U-NOT map and of the {\it optimal transpose} maps. The
measurement of the lower eigenvalue $\Lambda _{\min }$ of $\widetilde{{\Bbb I%
}_{A}\otimes K_{B}}(\rho _{AB})$ is a syndrome of the separability of the
state. In particular, $\Lambda _{\min }\leq \frac{2}{9}$ is found to imply
entanglement\ \cite{Horo02}. In this framework it is important to achieve a
high fidelity and reliable implementation of the stochastic optimal
transpose map.

Let us consider the most general single qubit map based on the $4-dim$
vectorial representation of the qubit density operator $\rho =%
{\frac12}%
({\Bbb I}+\overrightarrow{r}\cdot \overrightarrow{\sigma })$ \ written in
terms of the vector $(1,\overrightarrow{r})$ in the 4-dimensional space $%
\{\sigma _{0}\equiv {\Bbb I}{\bf ,}\sigma _{i},i=1,2,3\}$. Any map ${\cal E}$
is fully characterized by a $4\times 4$ real matrix ${\bf M}$,{\bf \ }which
maps $\rho $ into the density matrix: $\rho ^{\prime }={\bf M}\rho $. In
particular any {\it complete positive} map has the following matrix
representation 
\begin{equation}
{\bf M=}\left( 
\begin{array}{cc}
1 & 0 \\ 
\overrightarrow{t} & {\bf T}
\end{array}
\right)
\end{equation}
where ${\bf T}$ is a $3\times 3$ matrix and $\overrightarrow{t}$ is a $3-dim$
vector \cite{Oi01}. The $4\times 4$ matrix ${\bf M}_{TR}$\ associated with
the map ${\cal E}_{TR}$ transpose reads 
\begin{equation}
{\bf M}_{TR}=\left( 
\begin{array}{cccc}
1 & 0 & 0 & 0 \\ 
0 & 1/3 & 0 & 0 \\ 
0 & 0 & -1/3 & 0 \\ 
0 & 0 & 0 & 1/3
\end{array}
\right)  \label{matrixtranspose}
\end{equation}

We have implemented the optimal transpose gate by stochastically applying
the identity ${\Bbb I}$ and Pauli operators $\sigma _{X}$ and $\sigma _{Z}$.
In particular the random feature of the map was realized in an ergodic
fashion in the time domain, by alternate on/off switching of suitable
optical devices, indeed wp's, as shown shortly.

In the experiment carried out, the input qubit was codified into the $\pi
-state$ of a single photon and the Pauli operators $\sigma _{Z}$ and $\sigma
_{X}$ were realized adopting $\lambda /2$ waveplates respectively with angle
setting $\theta $ equal to $0^{\circ }$ and $45^{\circ }$. A very general
reconstruction of the optimal transpose transformation was experimentally
attained adopting the well known {\it Entanglement Assisted Quantum Process
Tomography }({\it EAQPT})${\it \ }$\cite{DeMa03}. This technique exploits
the quantum parallelism associated with any entanglement process: the
unknown map to be characterized acts on a subsystem of a bipartite entangled
state and all the information about the map is obtained from the
reconstruction of the{\it \ transformed} bipartite state. Indeed there is a
one-to-one correspondence between the map and the final state. In this way
only one input bipartite entangled state is needed to realize {\it EAQPT}.

Two entangled photons over the modes $k_{A}$ and $k_{B}$ in a {\it %
singlet-state} of polarization with common wl $\lambda =2\lambda _{p}=795nm$
were created by SPDC\ in a $1.5mm$ thick BBO NL\ crystal pumped by a
mode-locked beam with wl $\lambda _{p}$:\ (Fig. 7-{\bf a}). We have fully
characterized the implemented stochastic map ${\cal E}_{EXP}$ by
reconstructing the associated representation matrix ${\bf M}_{EXP}$ by using
of the input {\it entangled state} $\rho _{AB}$ of the two photons. The
first step consisted of performing the quantum state tomography of the input
system $\rho _{AB}$. In a following step\ the qubit associated with mode $%
k_{A}$, call it ''{\it qubit A}'', was left unchanged while the ''{\it qubit
B}'', associated with the mode $k_{B}$ underwent the ${\cal E}_{EXP}{\cal %
(\rho )}{\cal -}$transformation. Note that this procedure implies the
investigation of the unknown map ${\cal E}_{EXP}{\cal (\rho )}${\it ,} by a 
{\it complete} span over the Hilbert space ${\cal H}_{B}$ of the injected
''qubit $B$'' because of its fully {\it mixed-state} condition. The final
state of the two qubits $\rho _{AB}^{\prime }={\Bbb I}_{A}\otimes {\cal E}%
_{EXP-B}(\rho _{AB})$ was again investigated by tomographic
characterization. Finally, the matrix ${\bf M}_{EXP}$ was estimated by means
of the experimentally determined density matrices $\rho _{AB}$ and $\rho
_{AB}^{\prime }$ by adopting the relation$:{\bf M}_{EXP}^{T}={\bf C}^{-1}%
{\bf C}^{\prime }$. In this expression ${\bf C}_{i,j=0,3}=Tr[\left( \sigma
_{iA}\otimes \sigma _{jB}\right) \rho _{AB}]$ and ${\bf C}_{i,j=0,3}^{\prime
}=Tr[\left( \sigma _{iA}\otimes \sigma _{jB}\right) \rho _{AB}^{\prime }$
are the measured correlation matrices used to reconstruct $\rho _{AB}$ and $%
\rho _{AB}^{\prime }$ respectively$.$ We note that for an entangled state
the matrix ${\bf C}$ is always\ invertible. The {\it EAQPT} reconstruction
of ${\bf M}_{EXP}$ is shown in Fig.7-{\bf b. }

In order to compare the matrix $M_{EXP}$\ associated to the map ${\cal E}%
_{EXP}$ with the matrix $M_{TR}$ corresponding to the optimal transpose
operation ${\cal E}_{TR}$ (\ref{matrixtranspose}), we introduce the fidelity 
${\cal F}\left( {\cal E},{\cal L}\right) =\int d\Psi F\left[ {\cal E}\left(
\left| \Psi \right\rangle \left\langle \Psi \right| \right) ,{\cal L}\left(
\left| \Psi \right\rangle \left\langle \Psi \right| \right) \right] $ that
quantifies the overlap between two generic maps ${\cal E}$ and ${\cal L}$ $\ 
$\cite{Bowd02,DAri03}. In the present context we obtain: ${\cal F}\left( 
{\cal E}_{TR},{\cal E}_{EXP}\right) =1.01\pm 0.01$ . We may compare the
diagram reported in Fig.7-{\bf b} with the structure of the matrix ${\bf M}%
_{TR}$. The correspondence is quite impressive.

Note that the above stochastic transformation cannot be {\it reversed}, i.e.
the initial state can not be restored and the information encoded in the
output is lost in the environment. This is at variance with the optimal
Tele-UNOT Protocol described in Sections II and III \cite{Ricc03} or the
optimal UNOT gate based on the stimulated emission \cite{DeMa02}. These
transformations are indeed reversible: there the information about the input
qubit, redistributed into several qubits (the flipped qubit and the two
clones: ancilla qubits) can be reconstructed e.g. by a protocol suggested by
Bruss et al. \cite{Brus01}.

\section{Conclusions}

The Universal NOT gate and the Universal Optimal Quantum Cloning have been 
{\it contextually} implemented applying the projection over the symmetric
subspace to the input qubit and to an appropriate ancilla system. This
procedure has been found to consist of a modified quantum state
teleportation scheme. All these protocols, extended to the case of an
unlimited number of cloned or ancilla qubits, have been comprehensively and
straightforwardly accounted by a novel, very general approach based on the
well established angular momentum theory. By this approach many subtle
connections with the programmable optimal teleportation of other more exotic
anti-unitary transformations has been recovered. Most of these theoretical
results have been substantiated by the corresponding experiments, also
reported here. In particular, the linear (L)\ implementation of the {\it %
teleportation of a quantum gate} has been reported. It is an important tool
to be adopted for the realization of complex QI networks since it allows to
relax experimental constraints in order to achieve fault-tolerant processing 
\cite{Gott99}. Indeed the L-Optics quantum computation exploits the gate
teleportation in order to transform a probabilistic computation into a
nearly deterministic one \cite{Knil01}. Finally the stochastic feature of
the optimal partial transpose has been experimentally characterized in the
paper.

At last, it would be enlightening to compare the above results reported in
this paper with the ones obtained recently by the adoption of the {\it %
quantum injected} nonlinear (NL) parametric amplifier (QIOPA) \cite
{DeMa98,DeMa02,Lama02,Pell03}. There the symmetrization procedure implied by
cloning is provided automatically by the QED\ stimulated photon
amplification process involving at the same time the injected qubit and the 
{\it vacuum field}. In other words, the QIOPA realizes symmetrization,
cloning and entanglement within a unique fundamental, state-symmetrizing QED
process. Furthermore, there the vacuum amplification may be thought to
somewhat replace the mixed field associated with the mode $k_{A}$ in the
linear (L)\ symmetrization scheme shown in Fig. 4. In addition, the vacuum
amplification provides the squeezed vacuum noise (SVN)\ that necessary
affects in the QIOPA\ the {\it deterministic} realization of the non CP
(cloning and UNOT)\ maps. In the L case, the {\it non-probabilistic} \
optimal realization of these maps\ replaces {\it exactly} the amount of lost
information implied by SVN. In summary, the two conceptual approaches to
cloning already discussed in Section I, i.e. the symmetrization and the QED\
amplification, appear to be connected by subtle quantum mechanical links. We
believe to have enligthened in the present paper at least some of the most
interesting of these links. We believe that the actual results, the
suggestions and the open problems contributed by the present work could be
useful at least by setting measurement bounds and fundamental performance
limitations in the domain of Quantum Information and Quantum Estimation.
This work has been supported by the FET European Network on Quantum
Information and Communication (Contract IST-2000-29681: ATESIT), by Istituto
Nazionale per la Fisica della Materia (PRA\ ''CLON'')\ and by Ministero
dell'Istruzione, dell'Universit\`{a} e della Ricerca (COFIN 2002). F.S.
acknowledges ''Progetto Giovani Ricercatori'' (M.I.U.R.) for financial
contribution.

\centerline{\bf Figure Captions}

\vskip 8mm

\parindent=0pt

\parskip=3mm

Figure.1 General scheme for the simultaneous realization of the
Teleportation of the UNOT\ gate ($Tele-UNOT$) and the Universal Quantum
Cloning Machine (UOQCM) by applying a projective operator. The optional
insertion on the channel $A$ of a suitable unitary operator $U$ allows the
optimal teleportation of any anti-unitary map at Bob's site.

Figure.2. Realization of the UOQCM\ and of the Tele-UNOT gate by means of a
quantum circuit.

Figure.3. General scheme for the simultaneous realization of the $%
N\rightarrow (M-N)$ Tele-UNOT\ gate ($Tele-UNOT$) and of the $N\rightarrow M$
UOQCM by applying the state-symmetrization projective operator.

Figure.4 Setup for the optical implementation of the {\it Ou-Mandel Cloning }%
and corresponding results for three input qubits. {\it Filled squares:}
plots corresponding to the ''correct'' polarization; {\it Open triangles}:
plots corresponding to the ''wrong'' polarization. The solid line represents
the best gaussian fit expressing the {\it correct} polarization. These
options also apply to Figures 5 and 6.

Figure.5. Setup for the optical implementation of the{\it \ Cloning process }%
in a modified teleportation scheme. The corresponding results for three
input qubits are also reported.

Figure.6. Setup for the optical implementation of the Tele-UNOT gate and
corresponding results for three input qubits.\ 

Figure.7. Setup for the optical implementation of the Entanglement Assisted
Quantum Process Tomography ({\it EAQPT})\ of a stochastic map implementing
the optimal partial transpose {\bf (a)}. Experimental reconstructed matrix
operation ${\bf M}_{EXP}$ {\bf (b)}.

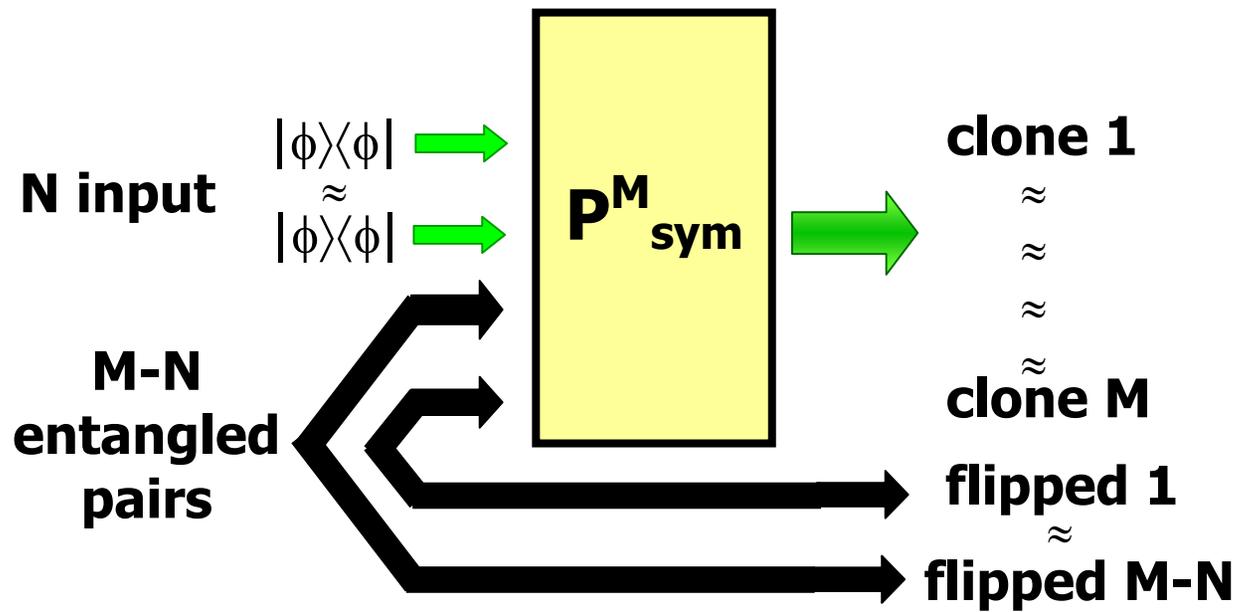
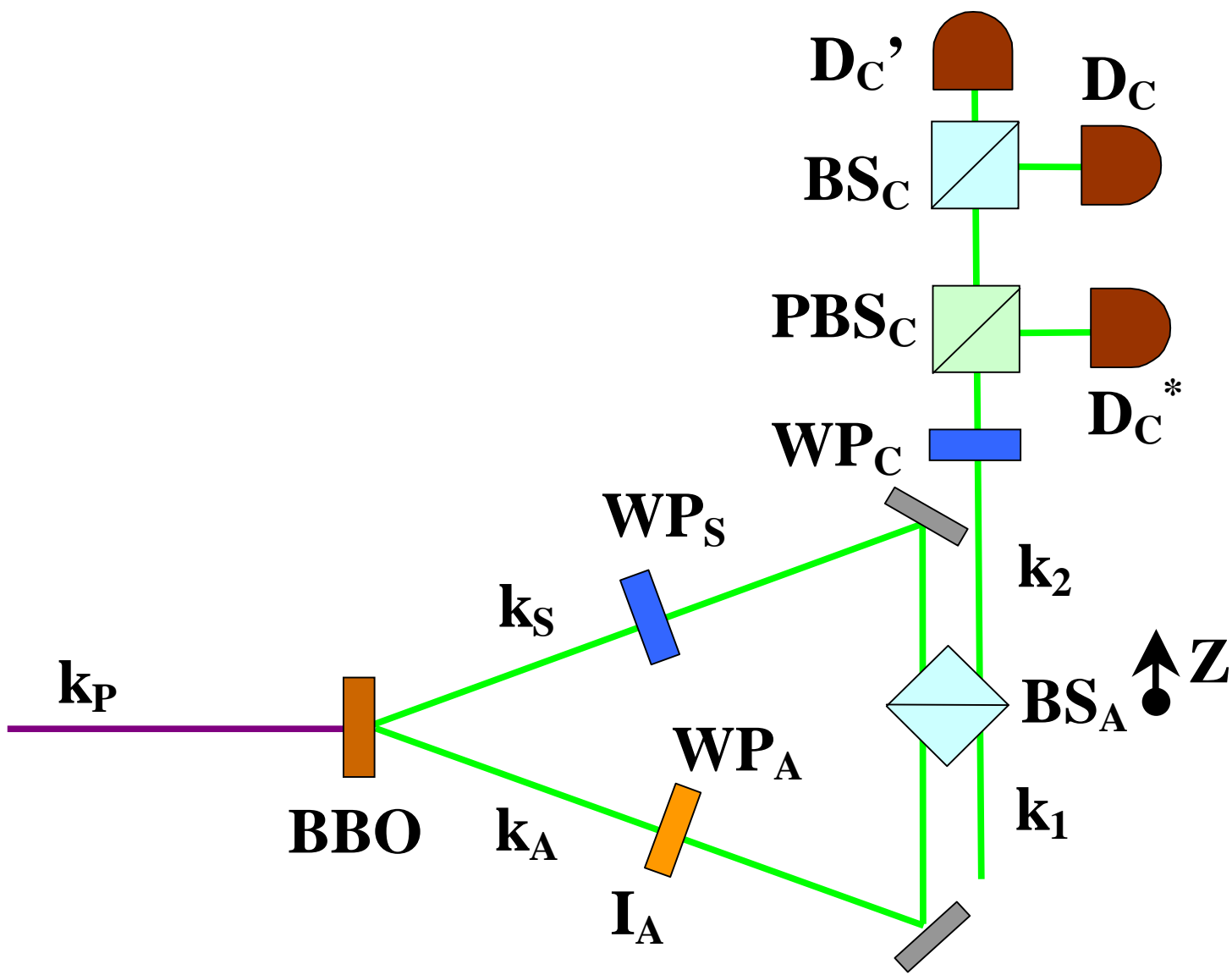
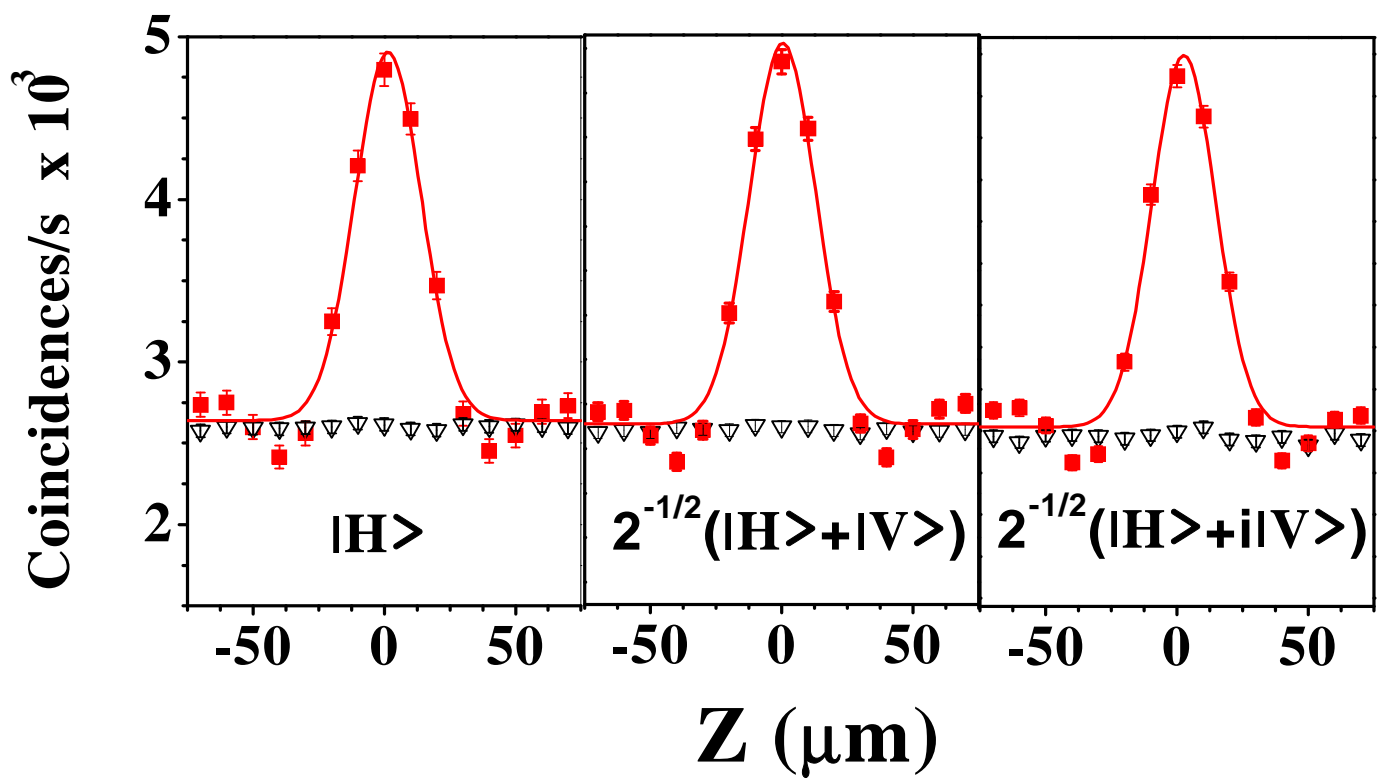
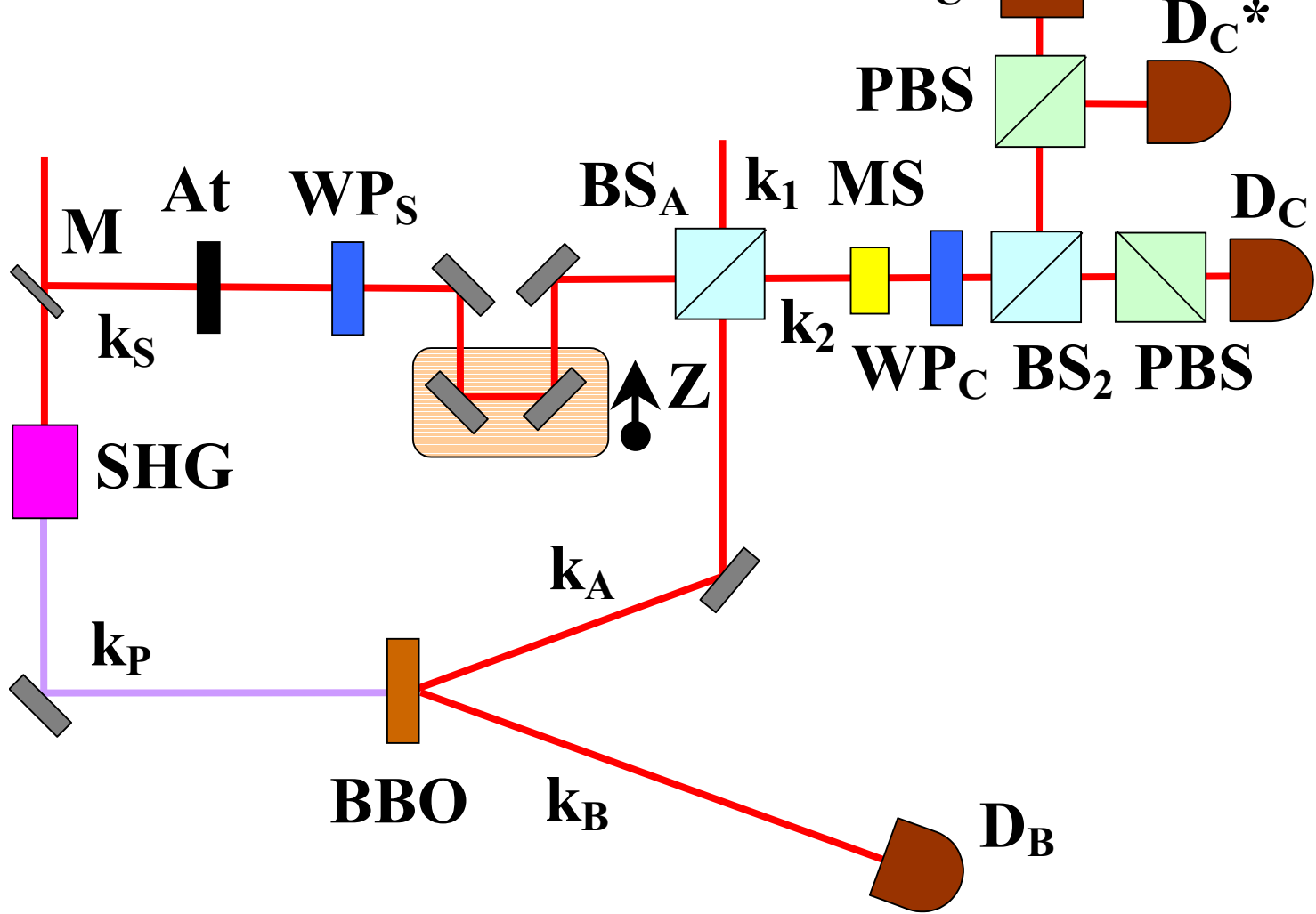
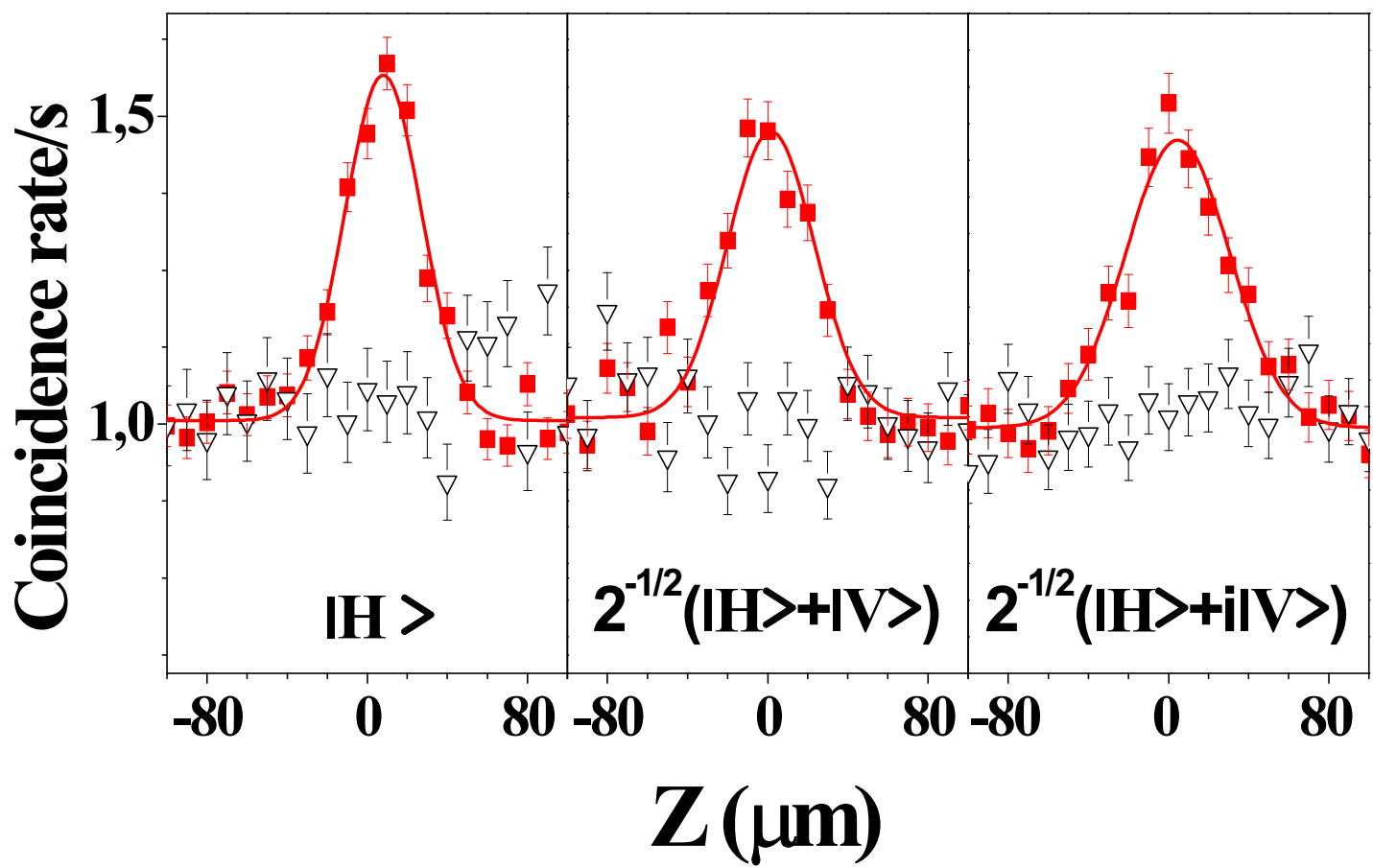
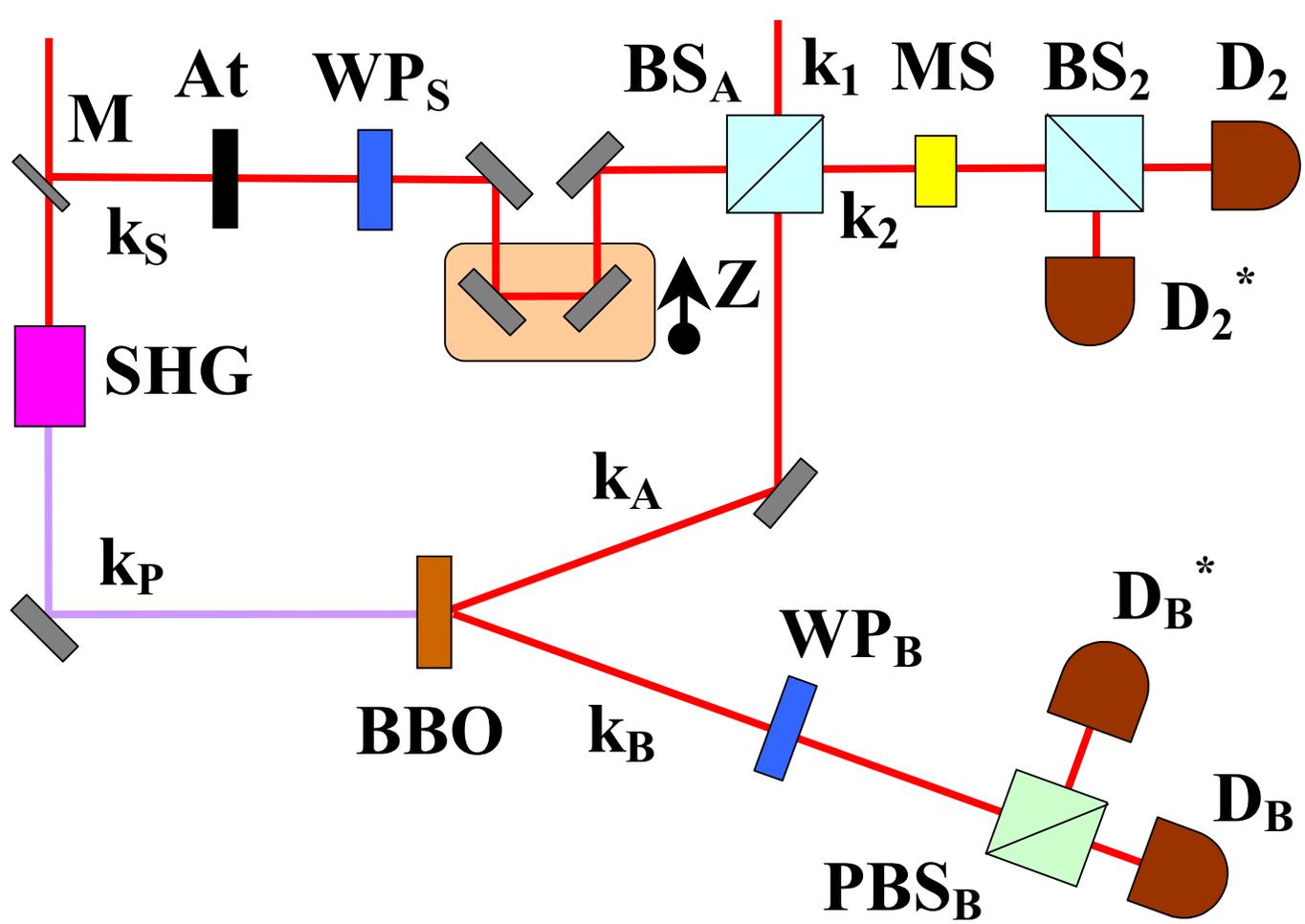
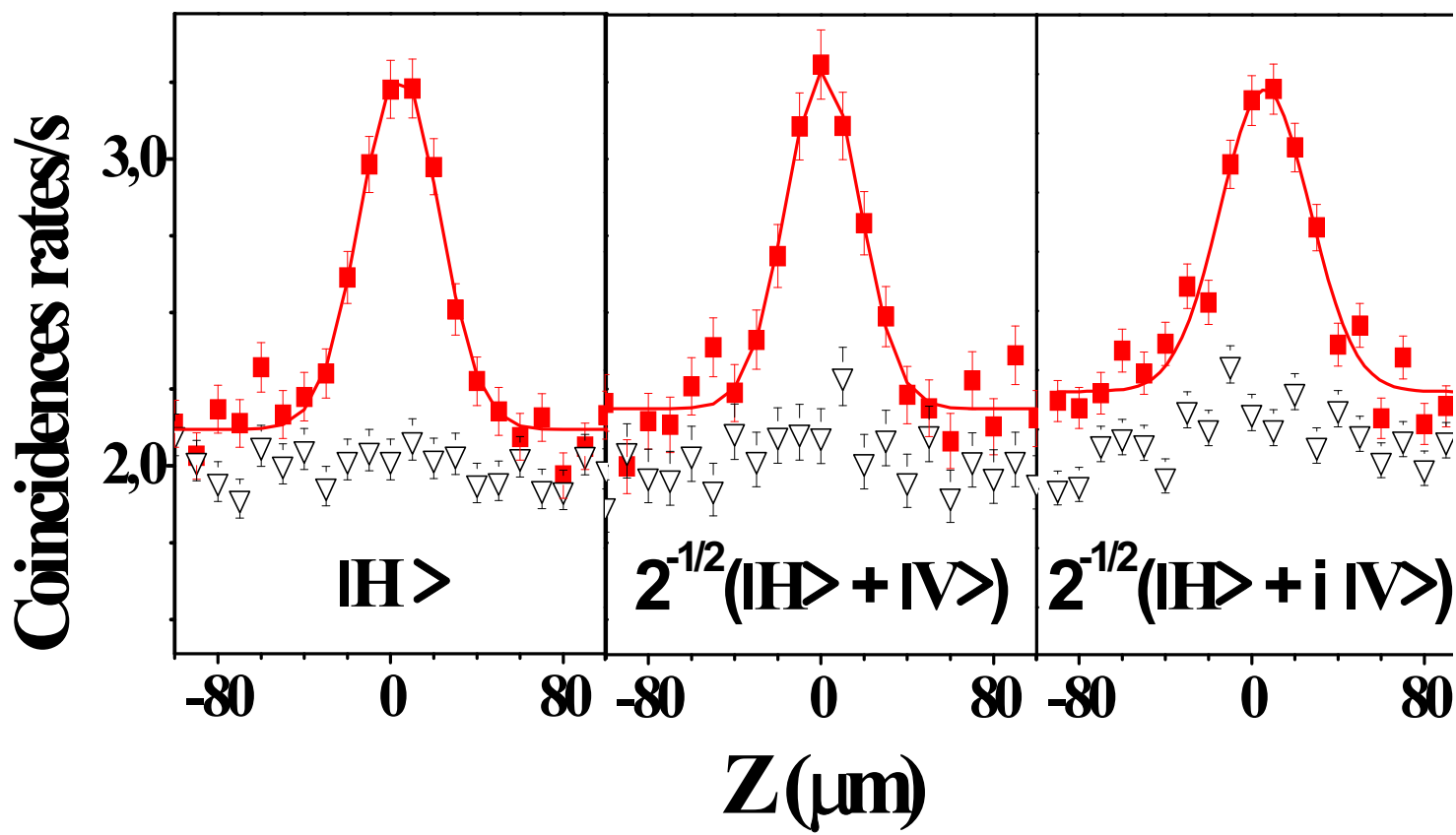
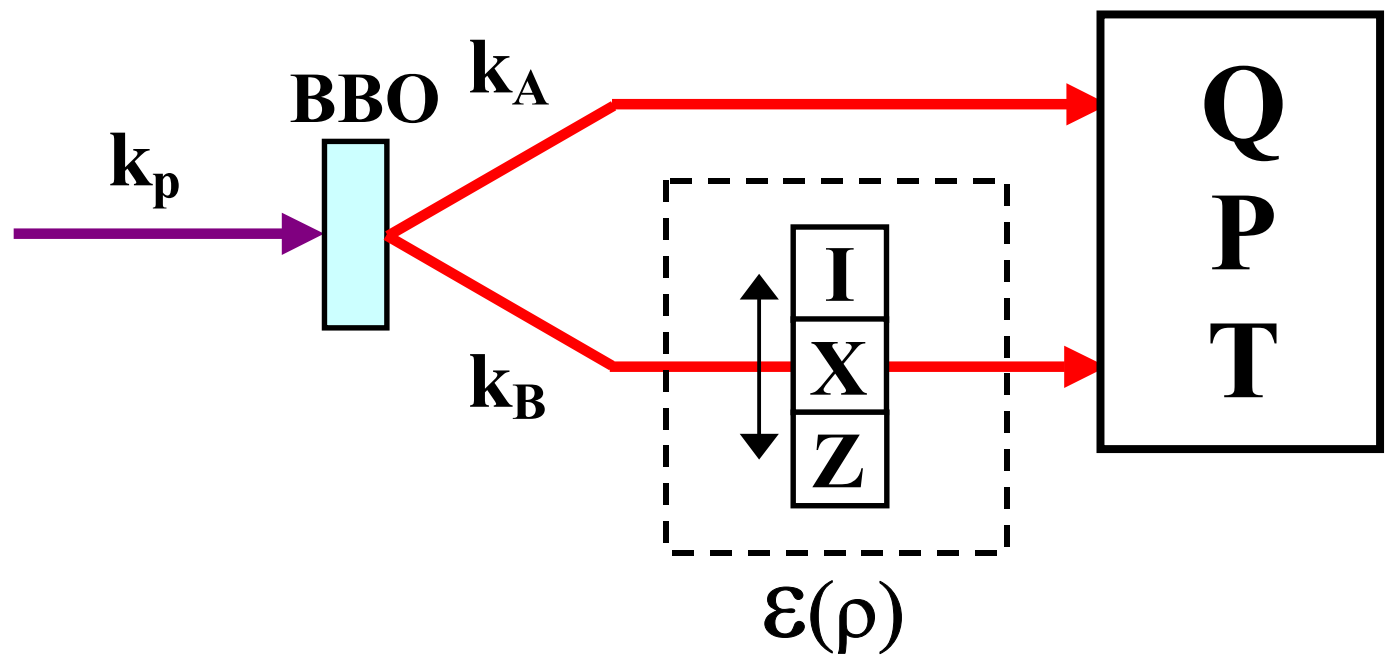
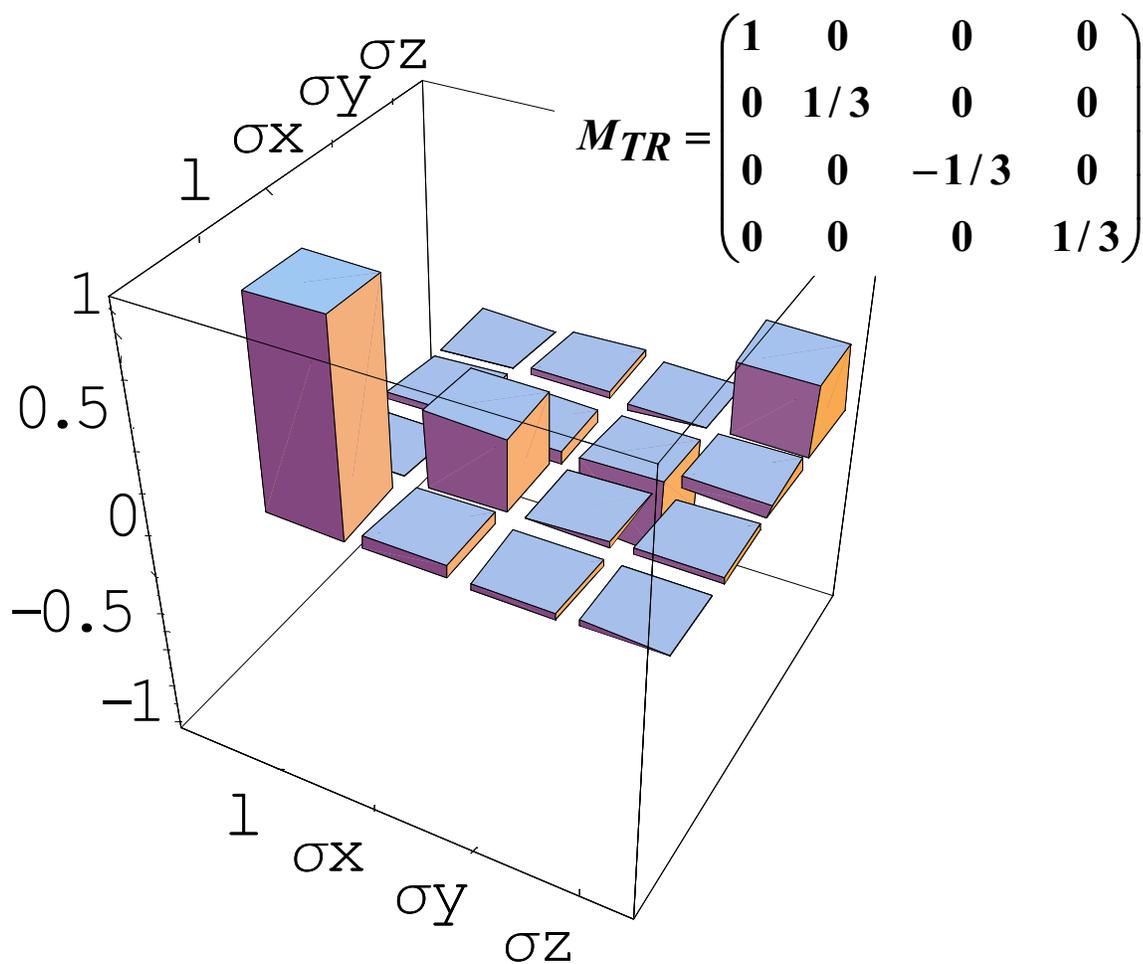
\end{document}